\documentclass[twocolumn,showpacs,amsmath,amssymb,amsfonts,nofootinbib,prl]
{revtex4-1}
\usepackage{graphicx}
\usepackage{dcolumn}
\usepackage{bm}
\usepackage{ulem}
\usepackage{overpic}

\def\beq{\begin{equation}}
\def\eeq{\end{equation}}

\begin{document}
\input{epsf}

\title{Heavy quark energy loss and $\mathbf D$-Mesons in {\it RHIC} and {\it LHC} energies}
\author{Raktim Abir$^1$, Umme Jamil$^2$, Munshi G. Mustafa$^1$, and
Dinesh K. Srivastava$^3$}
\affiliation{$^1$Theory Division, Saha Institute of Nuclear Physics  
1/AF Bidhannagar, Kolkata 700064, India.}

\affiliation{$^2$ Department of Physics, D. R. College, Golaghat, 
Assam 785621, India} 

\affiliation{$^3$Theory Group, Variable Energy Cyclotron Centre  
1/AF Bidhannagar, Kolkata 700064, India.}

\begin{abstract}
We obtain the radiative energy loss of a heavy quark in a deconfined
medium due to radiation of gluons off them using a recently derived 
generalised gluon emission spectrum. We find that the heavy flavour loses 
energy almost in a similar fashion like light quarks through this process. 
With this, we further analyse the nuclear modification factor for 
$D$-meson at LHC and RHIC energies. In particular, the obtained result
is found to be in close agreement with the most recent data from 
ALICE collaboration at 2.76 ATeV Pb-Pb collisions. We also discuss
the nuclear modification factor due to the collisional energy loss.
Furthermore, the result of non-photonic single electron from the decay of both 
$D$ and $B$ mesons is compared with the RHIC data at 200 AGeV Au-Au collisions, 
which is also in close agreement.

\end{abstract}


\date{\today}
\maketitle

\section {\bf Introduction}

The purpose of ongoing relativistic heavy ion collisions is to 
understand the properties of nuclear or hadronic matter at extreme conditions. 
A primary aim lies in the detection of a new state of matter formed 
in these collisions, the quark-gluon plasma (QGP), where the quarks and gluons 
are liberated from the nucleons and move freely over an extended region rather 
than over a limited hadronic volume. Various diagnostic measurements 
taken at CERN Super Proton Synchrotron (SPS)~\cite{heinz} in the past and at 
BNL Relativistic Heavy Ion Collider (RHIC)~\cite{white,dilep,phot,ellip,jet,
phenix,non-photonic} in the recent past have provided strong hints for 
the formation of QGP within a first few fm/$c$ of the collisions through the 
manifestation of hadronic final states.  New data from heavy-ion experiments 
at CERN Large Hadron Collider (LHC)~\cite{lhc,lhc1,lhc2} have further 
indicated the formation of such a state of matter.

One of the important features of the plasma produced in heavy-ion collisions 
is suppressed production of high energy hadrons compared to the case of pp
collisions, called jet quenching.
The term $`$jet quenching', generally, ascribes to the modification of 
an energetic parton due to its interaction with the coloured medium while 
passing through it. 
The basic idea is that the scales of hard (high-$p_\bot$) processes and 
the medium interactions in the context of heavy-ion collisions, are very 
distinct in accordance with the uncertainty principle.
This provides the fact that the high-$p_\bot$ parton production in $A-A$ 
collisions can be computed using perturbative QCD (pQCD), which is quite 
close to the vacuum rate scaled for binary $N-N$ collisions in an $A-A$ 
collision. The effect of medium is then treated as a final state interaction 
which is taken into account through the modification of the 
outgoing parton fragmentation pattern due to parton-medium interactions.      

The heavy-ion program at BNL RHIC~\cite{non-photonic} has clearly revealed 
that the phenomenon of jet quenching is mainly caused due to the energy loss 
of the initial hard parton via collisional and radiative processes, 
prior to hadronisation. The indication for jet quenching in heavy-ion program 
at CERN LHC has also been~\cite{lhc,lhc1,lhc2} observed recently.
The energy loss encountered by an energetic-parton in a QCD
medium reveals the dynamical properties of that medium and  
presently is a field of high interest in view of jet quenching of high 
energy partons; both light~\cite{bjorken82,thoma,mgm08,
gyulassy94,horowitz10, majum10,renk10,Baier:2000sb,zakharov96,wiedemann00,
gyulassy00,amy01,jeon05, gao00,salgado05,armesto,renk11}
and heavy quarks~\cite{horowitz10,braaten91,mgm05,peshier08,mgm97,
dokshit01,dead,dead1, Fochler:2010wn,alam10,kope10,Abir10,das10,
Fochler:2008ts,Gossiaux10,wicks07,jamil10,armesto1,armesto2,Uphoff11a,
Uphoff11b, Zhang04,xiang05,Vitev2008}. Naively, one imagines that the amount 
of quenching for heavy flavours jet should be smaller than that of light 
flavours due to the large mass of heavy quarks. However, the single electron 
data at RHIC~\cite{non-photonic} exhibit almost a similar suppression for heavy 
flavored hadrons compared to that for light hadrons.

A first attempt to estimate the radiative energy loss of heavy flavours 
in a QGP medium was made in Ref.~\cite{mgm97} by using the Gunion Bertsch 
(GB) formula~\cite{gunion82} of gluon emission for light quark scattering 
and appropriately modifying the relevant kinematics for heavy quarks.
Later the GB-like formula for heavy quarks was reconsidered in
Ref.~\cite{dokshit01} by introducing the mass in the matrix
element~\cite{dead} but only within the small angle approximation.   
Due to this mass effect, a suppression, known as $`$dead cone' effect,
in the soft gluon emission off a heavy quark was predicted in comparison 
to that from a light quark. This resulted in a reduction of heavy quark 
energy loss induced by the medium~\cite{dokshit01}, which is limited 
only to the forward direction.
However, such a gluon radiation spectrum with a dead cone factor, only 
applicable to the forward direction, was also used in the 
literature~\cite{dead1,alam10} uniformly for the full range of the 
emission angle ({\it i.e.}, both forward and backward direction) of 
gluon to calculate the heavy quark energy loss in the medium. This 
can lead to a unphysical result at large angle radiation, as discussed
as well as shown in Ref.~\cite{abir12}. Further attempts were also made 
in the literature~\cite{kope10,Gossiaux10,wicks07,armesto1,armesto2,
xiang05,Vitev2008} to improve the calculation of heavy quark energy 
loss with various ingredients as well as restrictions. In some 
cases
the energy loss for charm 
quark was found to be different than the light quark. The subject of
heavy quark energy loss is not yet a settled issue and requires more
detailed analysis. 

In a very recent work~\cite{abir12} the probability of gluon emission 
off a heavy quark has been generalised by relaxing some of the 
constraints, {\textit e.g.,} the gluon emission angle and the scaled 
mass of the heavy quark with its energy, which were
imposed in earlier calculations~\cite{dokshit01,dead}.
It resulted in a very compact and elegant expression for the
gluon radiation spectrum off a heavy quark ({\textit e.g.,} 
$Qq\rightarrow Qqg$) as~\cite{abir12} 
\begin{eqnarray}
\frac{dn_g}{d\eta d k_\bot^2}&=& \frac{C_A \alpha_s}{\pi}
\frac{1}{k_\bot^2}{\cal D} \ ,
\label{eq1}
\end{eqnarray}
where the transverse momentum of the emitted massless gluon is 
related to its
energy by $k_\bot=\omega\sin\theta$, and the rapidity,
$\eta=-\ln[\tan (\theta/2)]$,  is related to the emission angle,  
and the generalised dead cone is given by
\begin{eqnarray}
{\cal D} = \left(1+\frac{M^2}{s}e^{2\eta}\right)^{-2}  
=\left (1+\frac{M^2}{s\tan^2(\frac{\theta}{2})} \right )^{-2} \ .
\label{eq2} 
\end{eqnarray}
Now, the Mandelstam variable $s$ is given as, $s=2E^2+2E\sqrt{E^2-M^2}-M^2$, 
with $E$ and $M$, respectively, the energy and mass of the heavy quark. 
$C_A$ is the Casimir factor for adjoint representation and $\alpha_s$ is 
the strong coupling constant. In the small angle limit, 
$\theta \ll \theta_0(=M/E) \ll 1$, the dead cone in (2) reduces to that
in Ref.~\cite{dokshit01,dead} as $(1+\theta_0^2/\theta^2)^{-2}$ whereas for 
massless case it becomes unity and  (\ref{eq1})
reduces to the GB formula~\cite{gunion82}. The gluon spectrum for the 
process, $Qg\rightarrow Qgg$, can also be found in Ref.~\cite{abir12}.
We also note that the gluon emission spectrum in \eqref{eq1} is obtained 
in Feynman gauge. The same result is also obtained 
using light-cone gauge.

\begin{figure}
\includegraphics[width=0.8\linewidth, angle=0]{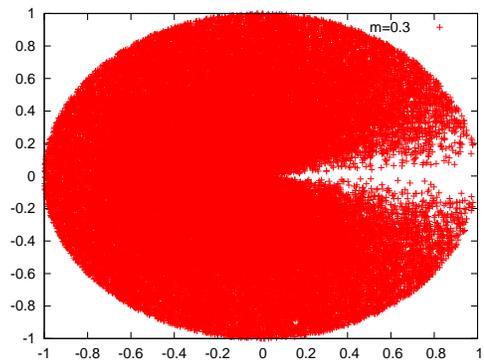}
\caption{(color online) A Monte Carlo simulation for the suppression factor
in \eqref{eq2} in the full domain of gluon emission angle, $\theta$, off a 
heavy quark for the scaled mass $m=\frac{M}{\sqrt s}=0.3$.
This actually represents a two dimensional view of the scaled gluon 
emission probability off a heavy quark with that of a light quark 
as given in \eqref{eq1}. We consider 
the direction of propagation of a heavy quark is from left to right 
along the horizontal axis and collide with medium partons at the 
origin of a circle of unit radius. This simulation has been performed 
by throwing points at random directions within the full domain 
of $\theta $ but with a probabilistic weight ${\cal D}(\theta)$, 
which would then correspond to a point randomly on the selected $\theta$-line 
as a $`${\textit{red plus}}' inside the circle of unit radius. 
The shade with red pluses represents 
the soft gluon emission zone whereas the conical white zone in the 
forward direction indicates a dead cone for gluon emission due to 
the mass effect.}
\label{simul}
\end{figure}

In Fig.~\ref{simul}, a Monte Carlo simulation of the above suppression 
factor (2) ({\textit i.e.}, the scaled gluon emission spectrum off 
a heavy quark with that of light quark) is displayed. It reveals a 
forward-backward asymmetry 
which encompasses the fact that the gluon emission off a heavy quark is 
as strong as that of light quark at the large angles (backward direction) 
whereas it is suppressed due to nonzero quark 
mass at the small angles (forward direction). However, if the energy of 
the heavy quark is large compared to its mass, the effect of dead cone
diminishes, both heavy and light quark are expected to lose energy
almost similarly. 
This result can have important consequences for a better understanding 
of heavy flavour energy loss in the context of heavy-ion collisions at RHIC 
and LHC. In this 
article we intend to use the gluon radiation spectrum in Ref.~\cite{abir12}
to obtain the heavy flavour energy loss and attempt to understand the 
suppression of heavy flavoured hadrons in heavy-ion collisions.
 
\section {\bf Radiative Energy Loss }
Among the interactions that a charged particle undergoes, as it traverses a 
dense matter, inelastic (i.e. radiative) scattering is undoubtedly the most 
important and interesting one. A number of different energy loss models has 
also been formulated in the literature (for review see 
Refs~\cite{armesto,renk11}). The basic differences among the different models
are the various constraints ({\it{e.g.}}, kinematic cuts, large angle 
radiation etc.) implemented to make the calculations manageable.
In this section we define the rate of radiative energy loss of a parton 
with energy $E$, due to inelastic scatterings with the medium partons
in a very canonical way as
\begin{eqnarray}
\frac{dE}{dx} &=&\frac{\langle \omega \rangle}{\langle \lambda \rangle},
\label{eq3}
\end{eqnarray}
where $\langle \omega \rangle$ and $\langle \lambda \rangle$ are 
the mean energy of emitted gluons 
and the mean free path of the traversing quark, respectively.

Among the set of variables $\left[k_\bot, 
\eta, \omega \right]$ in (\ref{eq1}) any two together are 
sufficient to completely describe an emitted gluon. For convenience we now 
change the variable duo from $[k_\bot, \eta]$ to 
$[\omega,\eta]$ as
\begin{eqnarray}
\frac{dn_g}{d\eta dk_\bot}&\Rightarrow &
\frac{dn_g}{d\eta d \omega} \ . \label{eq4}
\end{eqnarray}
It is now easy to find mean energy of the emitted soft gluons from the 
spectrum as
\begin{eqnarray}
\langle \omega \rangle &=&\left(\int~\frac{dn_g}{d\eta d \omega}
\omega~d\eta d\omega\right)/
\left(\int~\frac{dn_g}{d\eta d \omega}~d\eta d\omega\right)  \nonumber \\
&=&\left(\int~d\omega~\int \ {\cal D} d\eta\right)/
\left(\int~\frac{1}{\omega}~d\omega~\int {\cal D}\ d\eta\right) \ .
\label{eq5}
\end{eqnarray}
Other important quantity in $\eqref{eq3}$ is the  mean free path
$\langle \lambda\rangle$, which is 
the average distance covered by the traversing quark between two 
successive collision, {\it followed by a soft gluon radiation}. The magnitude 
of mean free path depends on the characteristics of the 
system in which the energetic particle is traversing, and 
it is defined as
\begin{eqnarray}
\langle \lambda \rangle &= & 1/(\sigma_{2\rightarrow 3}~\rho_{\rm qgp}) \ ,
\label{eq6}
\end{eqnarray}
where $\sigma_{2\rightarrow 3}\rho_{\rm qgp}=\rho_q\sigma_{Qq(\bar q)
\rightarrow Qq(\bar q)g} +\rho_g\sigma_{Qg\rightarrow Qgg}$, 
$\sigma_{2\rightarrow 3}$ is the  cross section of  
relevant $2 \rightarrow 3$ processes and $\rho_{\rm qgp}$ 
is the density of QGP medium which acts as a background containing 
target partons, for the high energetic projectile quark. 
We also note that the Landau-Pomeranchuk-Migdal (LPM) interference 
correction may be marginal, which we would estimate below based on the 
formation time of the emitted gluon along with the kinematical restrictions. 
Now, we recall the total cross section 
for $2\rightarrow 3$ processes as given in Ref.~\cite{Biro1993} as 
\begin{eqnarray}
\sigma_{2\rightarrow 3} & = & 2~C_A~{\alpha_s^3}~\int 
\frac{1}{(q_\bot^2)^{2}}~dq_\bot^2~\int~\frac{1}{k_\bot^2}~dk_\bot^2
\int {\cal D} \ d\eta \nonumber  \\
&= & 4~C_A~{\alpha_s^3}~\int \frac{1}{(q_\bot^2)^{2}}~dq_\bot^2~\int
\frac{1}{\omega}~d\omega~\int {\cal D} \ d\eta \ ,
\label{eq7}
\end{eqnarray}
where $q_\bot$ is the transverse momentum of the exchanged gluon.
Combining \eqref{eq6} and \eqref{eq7} the energy loss in \eqref{eq3} can
be written as
\begin{eqnarray}
\frac{dE}{dx} &=&12~{\alpha_s^3}~\rho_{\rm qgp}~\int_{\left.q_\bot^2
\right|_{\it min}}^{\left.q_\bot^2\right|_{\it max}}~\frac{1}{(q_\bot^2)^{2}}
dq_\bot^2 \nonumber \\
&&~~~~~~~~~~~~~\int_{\omega_{\it min}}^{\omega_{\it max}}~d\omega
\ 2 \int_{\eta_{\it min}}^{\eta_{\it max}}~{\cal D}\ d\eta \ ,
\label{eq8}
\end{eqnarray}
where a factor of 2  has been introduced in $\eta$ integral to cover 
both upper and lower hemisphere. We note that for ${\cal D}=1$,
\eqref{eq8} becomes equivalent to the massless case.

At this point it is important to note that the hierarchy employed in 
obtaining \eqref{eq1} in Ref.~\cite{abir12} reads as 
\begin{eqnarray}
{\sqrt s},E\gg \sqrt{|t|}\sim q_{\bot}\gg \omega > k_\bot \gg m_{D}\ ,
\label{eq9}
\end{eqnarray}
where $s$, $u$, $t$ are the usual Mandelstam variables 
and $m_{\tiny D}$ is the Debye screening mass of the thermal gluons.  
Based on the above hierarchy we obtain the kinematic cuts explicitly on
energy-momentum constraints and large angle radiation.
The infra-red cut-off has been used as
\begin{eqnarray}
\left.q_\bot^2\right|_{\it min} \simeq \omega^{2}_{\it min} 
\simeq \left.k_\bot^2\right|_{\it min} \simeq {m^{2}_{D}} 
= 4\pi \alpha_s T^2 \ . \label{eq10}
\end{eqnarray}
For ultraviolet cut-off on intermediate gluons, we have 
used~\cite{mgm97}, 
\begin{eqnarray}
\left.q_\bot^2\right|_{\it max} & = &  \frac{3}{2}ET-\frac{M^2}{4} \nonumber \\
&+& \frac{M^4}{48ET\beta_0}\log\left[\frac{M^2+6ET(1+\beta_0)}{M^2
+6ET(1-\beta_0)}\right]  , \label{eq11}
\end{eqnarray}
where $\beta_0 = (1-{M^2}/{E^2})^{1/2}$ 
and $T$ is temperature of thermal background. 
The ultraviolet cut-off on energy for the emitted soft gluon 
has been taken as average momentum of the intermediate 
gluon line as~\cite{Wang1995},
\begin{eqnarray}
\omega^{2}_{\it max} \simeq \langle q_\bot^2 \rangle \ . \label{eq12}
\end{eqnarray}
Now, the  relation between $\omega$ and $k_\bot$, $\omega=k_\bot \cosh \eta$, 
can be used to obtain bound on $\eta$ from top, which eventually excludes all 
collinear singularities for massless case. Finite cut on $\omega$ and $k_\bot$ 
then leads to an inequality, 
\begin{eqnarray}
\cosh \eta > \omega_{\it max} / \left.k_\bot\right|_{\it min} \ ,
\label{eq13}
\end{eqnarray}
from which one can easily obtain the bound on $\eta$ as
\begin{eqnarray}
\left|\eta\right|&<&\log\left(\frac{\sqrt{{\langle q_\bot^2 \rangle}}}
{m_D}+\sqrt{\frac{{{\langle q_\bot^2 \rangle}}}{m^2_D}-1}\right). \label{eq14}
\end{eqnarray}
We are now in position to discuss the LPM effect which
is usually included through a step function $\theta(\tau_i-\tau_f)$
while evaluating the spectrum of the radiated gluon. It
basically implies that the formation 
time of the gluon, $\tau_f= \langle \omega\rangle/\langle k_\bot^2\rangle$  
must be smaller than the interaction time $\tau_i\sim \Lambda^{-1}_{\rm{QCD}}=0.49/T_C$.  
This on the other hand imposes a 
restriction on the phase space of the emitted gluon as 
$\langle \omega \rangle > 2\Lambda_{\rm{QCD}} \approx 4T_C \sim gT 
\sim \mu_D$, provided $\alpha_s \sim 0.3$, $T_C\sim 170$ MeV
and the temperature of the plasma, $T\sim 350$ MeV. Thus, the hierarchy
in Eq.\eqref{eq9} excludes the modification of the radiative energy loss due to
the LPM interference correction through the infrared regulator, $\mu_D$. 
Therefore, the present formalism becomes akin to the Bethe-Heitler 
approximation, in which the scattering centers are well separated and 
the intensity of the induced radiation from different scatterings is additive. 

Now, it is very straightforward to obtain the radiative energy-loss
through the inelastic processes,
{\it viz.}, $Qq(\bar q)\rightarrow Qq (\bar q)g$
 and $Qg\rightarrow Qgg$ \cite{abir12},
for a heavy quark from \eqref{eq8}, which reads as 
\begin{eqnarray}
\left.\frac{dE}{dx}\right. &=&  24~{\alpha_s^3}~\left(\rho_q+\frac{9}{4}
\rho_g\right)~\frac{1}{\mu_g}~(1-\beta_1) \nonumber \\
&&\left(\frac{1}{\sqrt{(1-\beta_1)}}\left[\log\left(\beta_1\right)^{-1}
\right]^{1/2}-1\right)~{\cal F(\delta)}, \label{eq15}
\end{eqnarray}
where
\begin{eqnarray}
\!\!\!\!\!\!\! {\cal F(\delta)} &=& 2\delta-\frac{1}{2}\log\left(\frac
{1+M^2e^{2\delta}/s}
{1+M^2e^{-2\delta}/s}\right)  \nonumber \\
&-& \frac{M^2\cosh \delta/s}{1+2M^2\cosh \delta/s+M^4/s^2} \ , \nonumber \\ 
\delta &=&\frac{1}{2}\log\left[\frac{\log\beta_{1}^{-1}}{(1-\beta_1)}
\left(1+\sqrt{1-\frac{(1-\beta_1)^{\frac{1}{2}}}{\left[\log\beta_{1}^{-1
}\right]^{\frac{1}{2}}}}\right)^2\right] \ , \nonumber \\
s&=& E^2\left(1+\beta_0\right)^2 \ , \ \ \ \
\beta_1 = \frac{g^2}{\cal C}\frac{T}{E}\  , \nonumber \\
{\cal C} & = &  \frac{3}{2}-\frac{M^2}{4ET}\  \nonumber \\ 
&+& \frac{M^4}{48E^2T^2\beta_0}\log\left[\frac{M^2+6ET(1+\beta_0)}{M^2+6ET
(1-\beta_0)}\right] \ . \label{eq16}
\end{eqnarray}
Equation \eqref{eq15} together with \eqref{eq16} represents radiative energy 
loss of an energetic  quark in a canonical way within the framework 
of perturbative QCD along with kinematical restrictions for an energetic 
parton and medium interaction.

\section{Heavy quark production in pp collisions}

At leading order pQCD, heavy quarks in pp collisions are mainly produced by
fusion of gluons ($gg \rightarrow Q\overline{Q}$) or light quarks
($q \overline{q} \rightarrow Q\overline{Q}$)~\cite{comb}.
The cross-section  for the production of heavy quarks from pp collisions at
leading order can be expressed as~\cite{comb,cs1}:

\begin{eqnarray}
\frac{d \sigma}{dy_1\,dy_2\,dp_T}&=&2 x_1x_2 p_T\sum\limits_{ij}
\left [f_i^{(1)}(x_1, Q^2)f_j^{(2)} (x_2, Q^2)
\hat\sigma_{ij}+\right. \nonumber\\
&&\left. f_j^{(1)}(x_1, Q^2)f_i^{(2)}(x_2, Q^2)
\hat\sigma_{ij}\right ]/(1+\delta_{ij}),
\label{eq17}
\end{eqnarray}
where $i$ and $j$ are the interacting partons,
$f_i^{(1)}$ and $f_j^{(2)}$ are the partonic structure functions and
$x_1$ and $x_2$ are the fractional momenta of the interacting hadrons
carried by the partons $i$ and $j$.
The short range subprocesses for the heavy quark
production, $\hat\sigma\,=\,d\sigma/dt$ are defined as:
\begin{eqnarray}
\frac{d\sigma}{dt}=\frac{1}{16\pi {s}^2}\,|\mathcal{M}|^2,
\end{eqnarray}
where
 $|\mathcal{M}|^2$ for the processes $gg$$\rightarrow$$Q\bar{Q}$ and
 $q\bar{q}$$\rightarrow$ $Q\bar{Q}$ can be obtained from Ref.~\cite{comb}. 
The running coupling constant $\alpha_s$ at leading order is
\begin{eqnarray}
\alpha_s=
\frac{12\pi}{(33-2N_f)\ln \left(Q^2/\Lambda^2\right)},
\end{eqnarray}
where $N_f$\,=\,3 is the number of active flavours and
$\Lambda\,=\,\Lambda_{\rm QCD}$. The $p_T$ distribution 
of production of
heavy quarks at leading order supplemented with a K-factor $\approx$ 2.5 is
taken as the baseline for the calculation of the nuclear suppression factor,
$R_{AA}$~\cite{jamil10}. 
Effect of prefactor $K$ is diluted during computation of nuclear modification factor due to its identical 
effects on both initial and final distributions profiles. Furthermore, the K-factor, if equal for
$c$ and $b$ quarks, has not only a diluted effect but can actually be neglected in the ratios.
The shadowing effect is considered using EKS98 
parameterization~\cite{eks} for nucleon structure functions and here we 
use the CTEQ4M~\cite{cteq} set for nucleon structure function.  
We use Peterson fragmentation function with parameter $\epsilon_c=0.06$ and $\epsilon_b=0.006$ 
for fragmentation of $c$ quarks into $D$ mesons and $b$ quarks into $B$ mesons, respectively.

All the calculations are done assuming
the mean intrinsic transverse momentum of the partons to be zero.

\section{Initial Conditions and Evolution of the Medium}

As the heavy quarks are expected to lose most of their energy
during the earliest time after the formation of QGP, we can safely neglect
the transverse expansion of the plasma while discussing the heavy quark
energy loss.

We consider a heavy quark, which is being produced at a point ($r$,
$\Phi$) in a central collision and moves at an angle $\phi$
with respect to $\hat{ {r}}$ in the transverse plane. If $R$ be the radius 
of the colliding nuclei, the path length
covered by the heavy quark would vary from $0$ to $2R$, before it
exits the QGP. The distance covered by the heavy quark inside
the plasma in a central collision, $L$, is given by ~\cite{l}:
\begin{eqnarray}
L(\phi,\, { r})=
\sqrt{{ R}^2\,-\,{ r}^2\sin^2 \phi}\,-\,{ r}\cos\phi.
\label{eq18}
\end{eqnarray}
We can estimate the average distance traveled by the heavy quarks in the
plasma as:
\begin{eqnarray}
\langle L \rangle=
\frac{\int\limits_0^{ R}{ r}\,{ dr}
\,\int\limits_0^{2\pi} L(\phi,\, { r}) T_{AA}({ r}, b=0)\,{ d\,}\phi}
{\int\limits_0^{ R}{ r}\,{ dr}
\,\int\limits_0^{2\pi} T_{AA}({ r}, b=0)\,{ d\,}\phi},
\label{eq20}
\end{eqnarray}
where $T_{AA}(r, b\,=\,0)$ is the nuclear overlap function.
We estimate $\langle {L} \rangle$ as 5.78 fm
for central Au+Au collisions and 6.14 fm for central Pb+Pb collisions.

The temperature of the plasma at a time $\tau$,
assuming a chemically equilibrated plasma can be expressed as~\cite{wicks07}
\begin{eqnarray}
T\,(\tau) = \left(\frac{\pi^2}{1.202}\, \frac{\rho\,(\tau)}
{(9\,N_f +16)}\right)^{\frac{1}{3}}, \label{eq21}
\end{eqnarray}
where the gluon density at time $\tau$ is given 
by~\cite{wicks07}:
\begin{eqnarray}
\rho_g\,(\tau) = \frac{1}{\pi\,{ R}^2\,\tau}\, \frac{dN_g}{dy}.
\end{eqnarray}
Here we consider only the gluon density as the heavy quarks
lose most of their energy in interaction with gluons. We also add
that the gluon multiplicity is taken as $3/2$ times the number
of charged hadrons  and the initial temperature is obtained 
using (\ref{eq21}), assuming an initial time.

We take $(\frac{dN_g}{dy})$\,$\approx$\,1125
for Au+Au collisions at 200 AGeV~\cite{mrhic},
$\approx$\,2855 for Pb+Pb collisions at 2.76
ATeV\cite{lhc1}
and $\approx$\,4050 for Pb+Pb collisions at 5.5 ATeV~\cite{mlhc}.
We assume that the heavy quark having rapidity in the
central region moves along the
 fluid of identical rapidity. This kind of approximation
has been used earlier in
literature~\cite{ramona,sean,sourav}.

We calculate the initial temperature of QGP formation $T_0$
at 200 ATeV as 400 MeV, at 2.76 ATeV as 525 MeV and at
5.5 ATeV as 590 MeV, assuming the initial time of QGP formation
as $\tau_0$\,=\,0.2 fm/$c$. The critical temperature $T_c$\ for
the existence of QGP is taken as $\approx$\,170 MeV. The time, by
which the plasma will reach the critical temperature, $\tau_c$ is found to be
\,$\approx$\, 2.627 fm/$c$ at 200 AGeV, 5.9038 fm/$c$ at 2.76 ATeV
 and 8.375 fm/$c$ at 5.5 ATeV, assuming Bjorken's
cooling law, $T^3\, \tau$\,=\,constant.

The average path length of the heavy quark inside the plasma is
calculated as follows.
The velocity $v_T$ of a heavy quark can be expressed as $p_\bot/m_T$,
where $m_T$ is the transverse mass. Thus, the
heavy quark would cross the plasma in a
time $\tau_L$ \,= \,$\langle { L} \rangle/v_T$. Now,
if $\tau_c$ \, $\geq$ \, $\tau_L$, the heavy quark would remain
 inside the QGP during the entire
period, $\tau_0$ to $\tau_L$. But if $\tau_c$ \, $\textless$ \, $\tau_L$,
it would remain inside QGP only while covering the distance
$v_T\,\times\, \tau_c$. Thus, we further approximate
 the expanding and cooling plasma with one at a
temperature of T at $\tau$\, =\, $\langle { L} \rangle_{\rm eff}/2$, where
$\langle { L} \rangle_{\rm eff}$\,=\,min
$\left[\langle { L} \rangle,\, v_T \,\times \,\tau_c \right]$
(see Ref.~\cite{wicks07}).

\begin{figure}
\includegraphics[width=0.8\linewidth, angle=270]{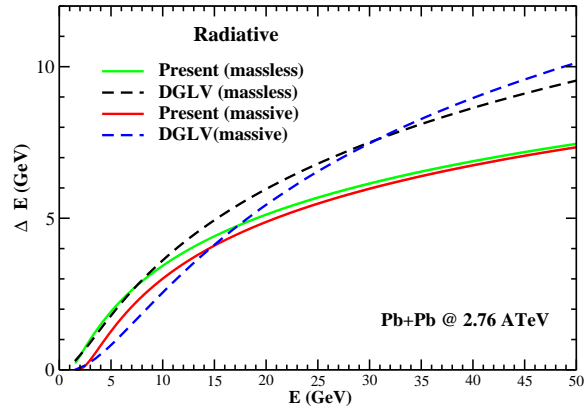}
\caption{(color online): Comparison of average energy loss for light quark 
and charm quark with mass $1.5$ GeV in a deconfined quark matter produced 
in Pb-Pb collision at 2.76 ATeV in the 
present and DGLV~\cite{wicks07} formalisms. For both cases the characteristics 
of the deconfined matter are treated in the same footing, {\textit i.e.}, the 
strong coupling $\alpha_s=0.3$ and the average path length, $\langle L \rangle 
\approx 6.14$ fm, traversed by an energetic quark in a deconfined medium
produced in such collisions.}
\label{fig1}
\end{figure}

\begin{figure}
\includegraphics[width=0.8\linewidth, angle=270]{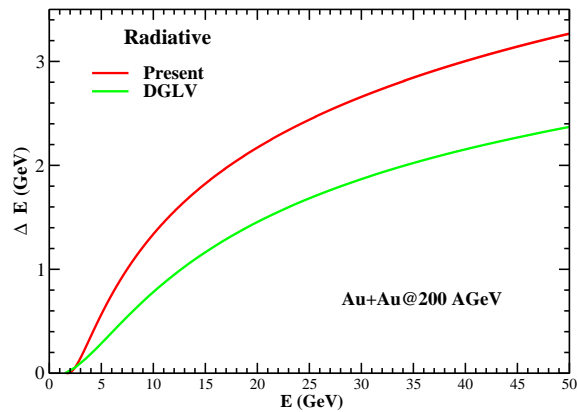}
\caption{(color online): Same as Fig.~\ref{fig1} but only for charm quark
in Au-Au collision at 200 AGeV with $\langle L \rangle =5.78$ fm.}
\label{fig2}
\end{figure}

\begin{figure}
\includegraphics[width=0.8\linewidth, angle=270]{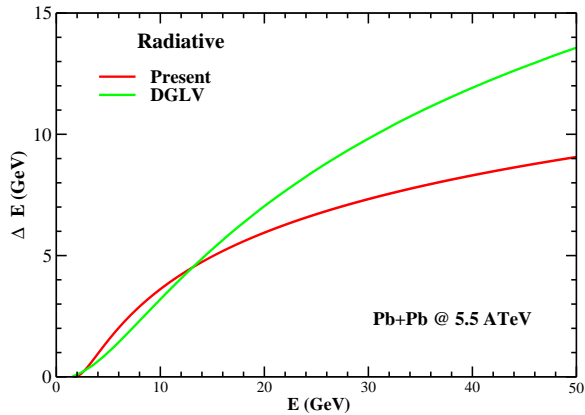}
\caption{(color online): Same as Fig.~\ref{fig2}
 in Pb-Pb collision at 5.5 ATeV with $\langle L \rangle =6.14$ fm.}
\label{fig4}
\end{figure}

\section{Results and Discussion}
In Fig.~\ref{fig1} a comparison of average radiative energy loss of 
an energetic quark traversing in a deconfined quark matter produced in Pb-Pb 
collision at 2.76A TeV in the present calculation with Djordjevic, Gyulassy,
Levai and Vitev  (DGLV) formalism in 
Refs.~\cite{gyulassy00,wicks07}.  As can be seen  both light 
and heavy quarks in the present formalism,  within the gluon emission 
spectrum of  ${\cal O}(\alpha_s)$ and ${\cal O}(1/k_\bot^2)$ as given in 
\eqref{eq1}, lose energy in a similar fashion for $E\geq 10$ GeV since the 
effect of mass is small compared to the energy. However, it is slightly 
less than that of a light quark for $E\leq 10$ GeV, due to the dead cone 
suppression at small angles. In addition the 
results from the present calculation differ from that of 
DGLV~\cite{gyulassy00,wicks07} one. These differences arise mainly because 
of the proper kinematic cuts for gluon emission as well 
as the method used to obtain energy loss. The various cuts in the 
present as well as in DGLV formalism are in close proximity except
the gluon emission in DGLV is constrained only to the forward emission 
angles~\cite{armesto}, 
$\theta \leq \pi/2$, whereas in the present calculation the full range
of $\theta$ is taken care off through the variable $\eta$ as shown in 
\eqref{eq13} and \eqref{eq14}.

In Figs.\ref{fig2}, and \ref{fig4} we have  displayed
average energy loss of a charm quark in a deconfined quark matter,
respectively, at 200 AGeV Au-Au collision at RHIC and 5.5 ATeV 
Pb-Pb collision at LHC. We find that at RHIC  energies the average energy loss 
of a charm quark in our formalism is higher than that of the DGLV formalism for 
the considered energy range, ($0 < E< 50$) GeV, of the charm quark. 
On the other hand  Fig.\ref{fig4} is 
qualitatively similar to  Fig.\ref{fig1} in terms of comparison of 
two formalism for heavy quark. As seen the average energy loss of 
charm quark is larger in the present formalism only in the domain, 
($0< E<15$) GeV, of the charm quark and beyond which it is less compared 
to the DGLV formalism.  The difference, in fact, increases as energy of 
the quark increases.

\begin{figure}
\includegraphics[width=0.8\linewidth, angle=270]
{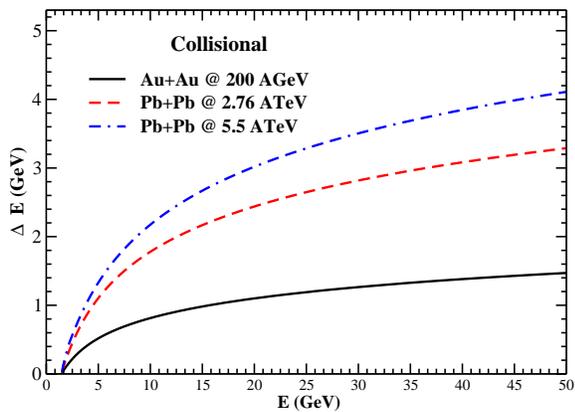}
\caption{(color online): Collisional energy loss of charm quark~\cite{peshier08} 
in Pb-Pb collision at 2.76 ATeV and 5.5 ATeV at LHC, and 200 AGeV
at RHIC energies.}
\label{coll_comp}
\end{figure}

\begin{figure}
\includegraphics[width=0.8\linewidth, angle=270]
{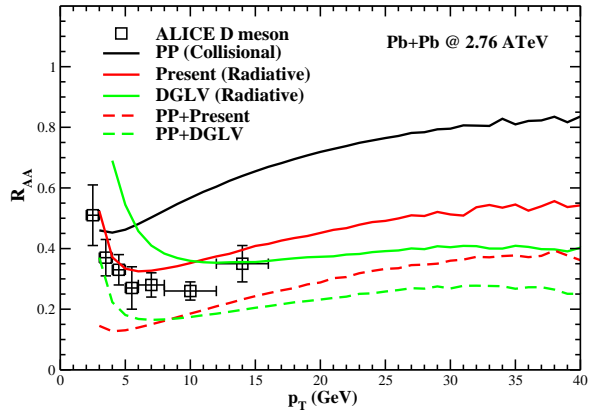}
\caption{(color online): Nuclear modification factor $R_{AA}$ for $D$ mesons 
with both collisional and radiative energy loss in Pb-Pb collision at 
2.76 ATeV. The data are from Ref.~\cite{exptl} but only the systematic error bars are
shown here.}
\label{Raa_rc}
\end{figure}

\begin{figure}
\includegraphics[width=0.8\linewidth, angle=270]{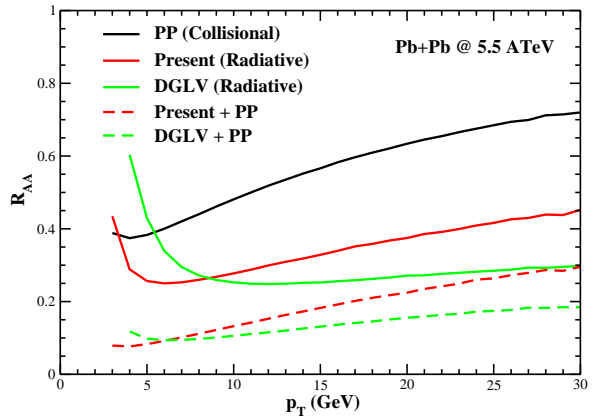}
\caption{(color online): Nuclear modification factor, $R_{AA}$,  for $D$ 
mesons in Pb-Pb collision at 5.5 ATeV.}
\label{Raa_3}
\end{figure}

In Fig.~\ref{coll_comp} we display a comparison of collisional energy loss
of charm quark as calculated by Peigne and Peshier (PP) in Ref.~\cite{peshier08} for 
RHIC and LHC energies. As seen the collisional energy loss increases with 
the increase in centre of mass energy of the colliding ions. 

In Fig.~\ref{Raa_rc} the nuclear suppression factor, $R_{AA}$, for $D$ meson
is displayed  considering both radiative and collisional energy loss and
compared with the ALICE data~\cite{exptl} at $2.76$ ATeV. As can be seen the 
differences in radiative energy loss between the present and DGLV formalism
discussed in Fig.~\ref{fig1} for $2.76$ ATeV in Pb-Pb collisions
is clearly reflected in Fig.~\ref{Raa_rc}. For the present calculation 
it is manifested in gradual increase of $R_{AA}$ of $D$ meson~\cite{exptl} 
for transverse momentum, $p_\bot > 5$ GeV whereas in DGLV case 
it remains almost constant. The suppression factor obtained in the present 
formalism with radiative energy loss is in close agreement with the 
most recent data from ALICE collaboration at 2.76 ATeV~\cite{exptl}.
On the other hand the inclusion of the collision contribution 
is found to suppress $R_{AA}$ further in both cases.  As found the data suggest
that the collisional contribution may be small. 
Nonetheless, more data in the high $p_\bot$ 
domain is necessary to know the actual trend of the energy loss of charm quark
and will finally constrain the various energy loss and jet quenching model 
in the literature.
We also expect a similar rise in light hadrons for high $p_\bot$ since both 
light and heavy quark lose energy in a similar fashion as shown in 
Fig.~\ref{fig1}.  However, we note that the ALICE data on $R_{AA}$ for 
inclusive charge hadrons~\cite{charged} at $2.76$ ATeV 
in Pb-Pb collision has also shown a similar increasing trend as $p_\bot$ 
increases. It is natural to believe that such data is completely dominated
by the contribution from light hadrons.  For completeness, we also display $R_{AA}$ 
for LHC energy at $5.5$ ATeV  in Fig.~\ref{Raa_3}.

\begin{figure}
\includegraphics[width=0.8\linewidth, angle=270]
{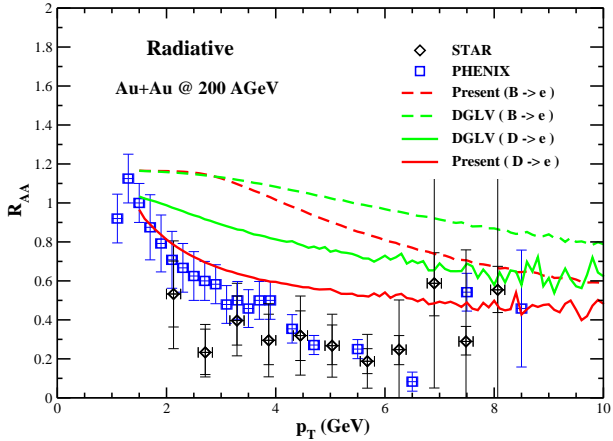}
\caption{(color online): $R_{AA}$ with only
radiative energy loss for non-photonic single 
electron from the decay of individual $D$ mesons and $B$ 
mesons in Au-Au collision at 200 AGeV. The data are from Ref.~\cite{non-photonic}.
Both systematic and statistical error bars are shown for STAR data whereas 
only systematic error bars are displayed for PHENIX data.}
\label{Raa_r_rhic}
\end{figure}

\begin{figure}
\includegraphics[width=0.8\linewidth, angle=270]
{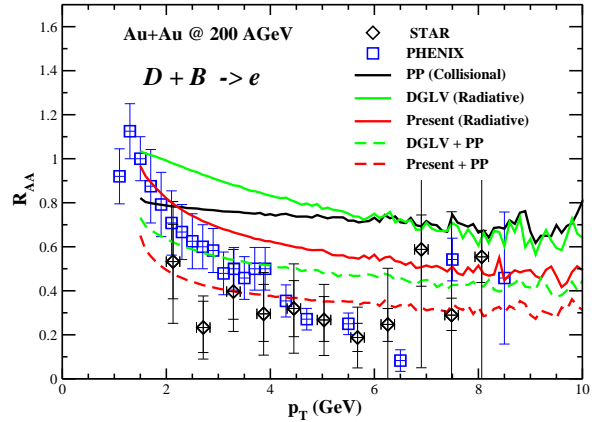}
\caption{(color online): $R_{AA}$ with collisional
and radiative energy-loss for non-photonic single 
electron from the combined decay of both $D$ and $B$ 
mesons in Au-Au collision at 200 AGeV.The data are from Ref.~\cite{non-photonic}. 
Both systematic and statistical error bars are shown for STAR data whereas only systematic error 
bars are displayed for PHENIX data.}
\label{Raa_rc_rhic}
\end{figure}

\begin{figure}
\includegraphics[width=0.8\linewidth, angle=270]{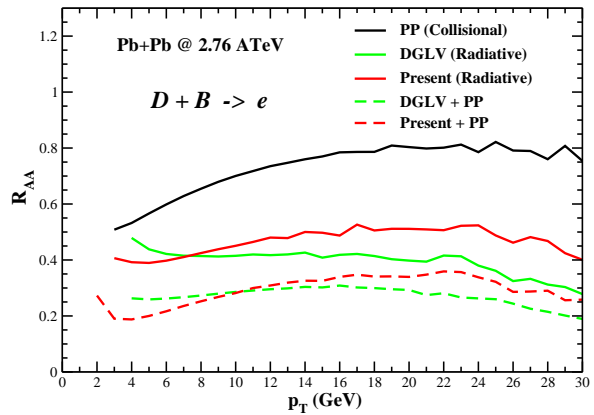}
\caption{(color online): Same as Fig.~{\ref{Raa_rc_rhic}}  in Pb-Pb collision at 2.76 ATeV.}
\label{Raa_4}
\end{figure}

In Fig.~\ref{Raa_r_rhic} the nuclear suppression factors for individual decay of $D$ and $B$ mesons 
to non-photonic single electron is displayed considering only the radiative energy loss for RHIC
energy at 200 AGeV. As expected the contribution from the $B$ decay is small compared to that of $D$ decay.
In Fig.~\ref{Raa_rc_rhic}  the total contribution of single electron from  $D$ and $B$ decay is shown 
considering both radiative and collisional energy loss. It is found that the contributions of the collisional energy 
loss is important at RHIC energy. We also compare our results with that
of DGLV. In Fig.~\ref{Raa_4}, we give prediction for single electron result for LHC energy at 2.76 ATeV.  

\section{Summary}

We obtain the radiative energy loss of a heavy quark akin to the Bethe-Heitler 
approximation by considering the most generalised gluon emission multiplicity
expression derived very recently. This suggests that both energetic heavy 
and light quark lose energy due to gluon emission almost similarly and 
the mass plays a role only when the energy of the quark is of the order of it.
The hierarchy used for simplifying the matrix element as well as  for obtaining
the gluon radiation spectrum imposes a restriction on the phase space of the
emitted gluon in which the formation time is estimated to be less than the 
interaction time. This suggests that the LPM interference correction may be 
marginal.  Further, we compare our results with the DGLV formalism and
it is found to differ significantly.  To compute the nuclear suppression 
factor for $D$-meson we consider both radiative and collision energy loss 
along with longitudinal expansion of the medium. The nuclear modification 
factor for $D$-meson with radiative energy loss obtained in the present 
formalism has an increasing trend at high $p_\bot$ and found to agree closely with 
the very recent data from ALICE collaboration at
2.76 ATeV. When the collisional counter part is added independently, the 
further suppression is obtained in the nuclear modification factor. 
This suggest  
The non-photonic single electron data at 200 AGeV RHIC energy requires
contributions from collisional energy loss as well from $B$ decay.
However, 
it is  necessary to obtain both radiative and collisional energy loss from 
the same formalism to minimize the various uncertainties, which is indeed a 
difficult task.  Moreover, data at high $p_\bot$ region with improved 
statistics  are required to remove prejudice on different
energy loss and jet quenching models.

\vspace*{0.3in}

\begin{acknowledgments}
RA gratefully acknowledges useful discussion with J. Uphoff
during the course of this work. UJ gratefully acknowledges the hospitality
at Saha Institute of Nuclear Physics where part of the work was done.
\end{acknowledgments}

\end{document}